\pdfoutput=1 

\documentclass[aps,pra,twocolumn,superscriptaddress,showpacs]{revtex4}

\usepackage{amssymb,amsmath,graphicx}

\begin{document}

\title{Conductor-Insulator Crossover in the Steady-State \\
       Ultracold Plasmas}

\author{Yurii V. Dumin}
\email[Electronic mail: ]{dumin@pks.mpg.de, dumin@yahoo.com}
\affiliation{Lomonosov Moscow State University,
Sternberg Astronomical Institute, \\
Universitetskii prosp.\ 13, 119234 Moscow, Russia}
\affiliation{Space Research Institute of Russian Academy of Sciences, \\
Profsoyuznaya str.\ 84/32, 117997 Moscow, Russia}
\author{Ludmila M. Svirskaya}
\email[Electronic mail: ]{svirskaialm@susu.ru, svirskayalm@mail.ru}
\affiliation{South Ural State University,
Prosp.\ Lenina 76, 454080 Chelyabinsk, Russia}
\affiliation{South Ural State Humanitarian Pedagogical University, \\
Prosp.\ Lenina 69, 454080 Chelyabinsk, Russia}

\date{May 20, 2026}

\begin{abstract}
We present a theoretical model of the ionization--recombination balance in
the ultracold Rydberg gas--plasma mixture, which is caused by the collective
processes rather than by individual interparticle interactions.
We consider the electron--ion system where each electron moves for the most
part of its time in the effective centrifugal potential formed by the nearest
ion and sometimes jumps to the neighboring potential wells due to
the perturbations from remote particles.
These perturbations are described by a thermal bath with the effective
``virial'' temperature.
Then, the electron with a sufficiently high energy, which can escape from
the local potential well, should be considered as free (conducting) one;
while the electron with low energy, confined within the well, as forming
the Rydberg atom.
As follows from our calculations, there is a sharp crossover from
the insulating phase (Rydberg gas) to the conducting one (plasma) with
increase in the particle density, which somewhat resembles the Mott
transition in condensed-matter physics.
This model should be well relevant to the steady-state ultracold plasmas
obtained in the most recent experiments and represents a promising direction
for further research.
\end{abstract}


\maketitle

\section{Introduction}
\label{sec:Intro}

The processes of mutual transformation from the dense Rydberg gas to plasma
and \textit{vice versa} were experimentally observed for a quite long time
both in the atomic beam experiments~\cite{Vitrant_82,Morrison_08} and
magneto-optical traps~\cite{Killian_01,Robert_13}.
From the theoretical point of view, they were commonly described as a sequence
of interactions between the individual particles, such as electron--Rydberg
and Rydberg--Rydberg collisions, three-body recombination, \textit{etc.}
with the respective cross-sections.

Meanwhile, as was emphasized already in the seminal paper~\cite{Vitrant_82}
in 1982, ``it is thus possible that the two-body analysis developed here is
too naive'', so that the effect of entire environment should be taken into
account.
This was qualitatively associated with Mott transitions in
the condensed-matter physics~\cite{Mott_74}, although the specific physical
mechanisms in these two cases are not the same.
Particularly, it was concluded in the above-cited paper that this situation
will take place if the atomic temperature is below a fraction of Kelvin.
Then, the first interatomic collisions---initiating a cascade of
the bi-particle interactions---will be delayed, and the collective processes
can be important.

Unfortunately, such cold atoms were unavailable in the early 1980s.
However, due to development of the magneto-optical traps, an even lower
temperature was achieved by the late 1990s, and the ionization--recombination
processes in the so-called ultracold plasmas (UCP) began to be studied in
these
installations~\cite{Killian_99,Gould_01,Bergeson_03,Killian_07a,Killian_07b}.
In fact, the most of such experiments were conducted in the strongly
non-equilibrium regime, when a bunch of more energetic electrons quickly
accelerated the ionic cloud, originally possessing the milliKelvin
temperatures.
Fortunately, the situation changed in the last couple years, when
the ultracold plasmas were generated in the steady-state
regime~\cite{Zelener_24}.
Thus, the prospect emerged for high-precision measurements of
the ionization--recombination processes---similar to those conducted in
the atomic beams---and, in particular, for studying the Mott-like
transitions.

It is the aim of our paper to present a model of collective development
of the ionization--recombination balance in the disordered non-degenerate
Rydberg gas and to discuss its observational properties.
As distinct from the previous works, \textit{e.g.}~\cite{Zelener_04},
treated in detail the population of individual Rydberg states
at the sufficiently early stage of the plasma cloud expansion (roughly
speaking, within 100--200\,$ \mu $s), we shall consider much longer times,
when the Rydberg states are completely mixed and their quasi-equilibrium
distribution is established.
In this sense, our approach is substantially different from
the theoretical models employed before.

\section{Theoretical Model}
\label{sec:Model}

\subsection{Basic Assumptions}
\label{sec:Assum}

\begin{figure}[b]
\includegraphics[width=0.6\columnwidth]{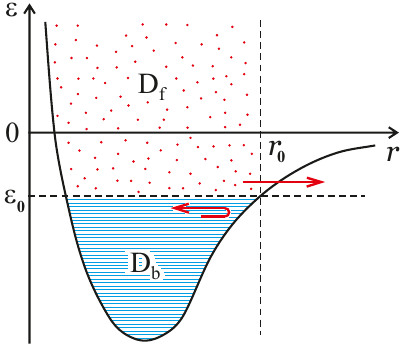}
\caption{\label{fig:Phase_space}
Phase space of an electron, composed of the regions of free and bound
states, $ {\rm D}_{\rm f} $ and $ {\rm D}_{\rm b} $ respectively.}
\end{figure}

We shall consider a system of almost immobile closely-packed Rydberg atoms
in the classical approximation.
(As is known, the purely classical approach is quite reasonable for
the description of Rydberg atoms~\cite{Gallagher_94,Lebedev_98}.)
Then,
in the strongly-coupled regime (\textit{i.e.}, when the electron coupling
parameter is about unity, $ {\Gamma}_e{\sim}1 $),
each electron should move for the most of time in the effective
``centrifugal'' potential~\citep{Landau_76} of the nearest ion,
\begin{equation}
U_{\rm eff}(r) =
  \frac{M_e^2}{2 m_e} \, \frac{1}{r^2} - \, e^2 \frac{1}{r} \, ,
\label{eq:Eff_pot}
\end{equation}
and sometimes jump to the neighbouring potential wells due to
the multi-particle perturbations from the environment.
Here,
$ M_e $~is the angular momentum of the electron with respect to
the corresponding ion,
$ m_e $ and $ e $~are the electron mass and charge, and
$ r $~is the distance between the electron and ion.
The corresponding curve~$ \varepsilon = U_{\rm eff}(r) $ is drawn in
Fig.~\ref{fig:Phase_space}.
It is important to emphasize that, according to the general theorem of
classical mechanics~\citep{Landau_76}, the angular momentum~$ M_e $ is
adiabatic invariant, \textit{e.g.}, approximately conserved for some time
even if the electron--ion pair experiences the small external perturbations.

Let $ r_0 $~be half the average distance between the ions (\textit{i.e.},
a characteristic size of the potential well) and $ {\varepsilon}_0 $~be
the corresponding value of the effective potential:
\begin{equation}
{\varepsilon}_0 = U_{\rm eff}(r_0) \, .
\label{eq:eps_0}
\end{equation}
Then, the electron phase space will be composed of two subregions,
$ {\rm D}_{\rm f} $ and $ {\rm D}_{\rm b} $.
Electrons in the region~$ {\rm D}_{\rm f} $, possessing the energy
$ \varepsilon\,{\geqslant}\,{\varepsilon}_0 $, can reach the distance~$ r_0 $
and, therefore, jump to a neighbouring potential well, as marked by the red
arrow in Fig.~\ref{fig:Phase_space}.
Consequently, they should be treated as free (which is denoted by
subscript~`f').
On the other hand, electrons in the region~$ {\rm D}_{\rm b} $ (with energy
$ \varepsilon\,{<}\,{\varepsilon}_0 $) cannot reach the distance~$ r_0 $ and,
therefore, will be reflected back from the potential wall.
So, they should be treated as bound, \textit{i.e.}, forming the Rydberg atoms.

If $ F(r, \varepsilon) $ is the partition function of the electron in
variables~$ r $ and~$ \varepsilon $, then concentrations of the free and
bound electrons can be written as
\begin{equation}
N_{\rm f} = \!
  \iint\limits_{{\rm D}_{\rm f}} \! F(r, \varepsilon) \, dr \, d\varepsilon
\;\; \mbox{and} \;\;
N_{\rm b} = \!
  \iint\limits_{{\rm D}_{\rm b}} \! F(r, \varepsilon) \, dr \, d\varepsilon
\, ,
\label{eq:Nf_and_Nb-gen_def}
\end{equation}
respectively.
This function~$ F $ can be constructed by the following way.

\subsection{Partition Function and Effective Temperature}
\label{sec:Partition}

In a reasonable approximation, the effective electron partition function in
the standard radius--velocity variables can be taken as Boltzmann
distribution,
\begin{equation}
f_{\rm eff}(r, v_r) =
  A_f \exp\big[{-}\varepsilon(r, v_r) / k_{\rm B} T_{\rm vir}\big] \, ,
\label{eq:f-gen_def}
\end{equation}
under assumption that the considered electron--ion pair is immersed into
a thermal bath with the effective (``virial'') temperature~$ T_{\rm vir} $
formed by all other charged particles, as will be explained below.
Here,
\begin{equation}
\varepsilon(r, v_r) = m_e v_r^2 / 2 \, + U_{\rm eff}(r)
\label{eq:total_energy}
\end{equation}
is the total energy of the electron,
$ v_r $~is its radial velocity,
$ k_{\rm B} $~is Boltzmann constant, and
$ A_f $~is the normalization factor.
Then, performing a transformation from the radius--velocity to radius--energy
variables by formula~(\ref{eq:total_energy}), we get the required partition
function in the form:
\begin{equation}
F(r, \varepsilon) =
  A_F \exp\Big\{{-}\frac{\varepsilon}{k_{\rm B} T_{\rm vir}} \Big\}
  \frac{1}{\sqrt{\varepsilon - U_{\rm eff}(r)}} \, ,
\label{eq:F-init}
\end{equation}
where $ U_{\rm eff} $ is given by formula~(\ref{eq:Eff_pot}).

As was already mentioned above, collective description of
the ionization--recombination processes should be appropriate in
the case of sufficiently low temperatures and, therefore, relatively small
electron velocities.
However, it should be kept in mind that in the multi-body Coulomb system
\textit{
the electron kinetic energy can be arbitrarily large as compared to the
potential energy, but it cannot be arbitrarily small.}
Really, if the kinetic energy dropped down to the value about the potential
one, the slow electrons are accelerated by the nearest ions and begin to fall
onto them, thereby performing a quasi-bound motion in the approximately
elliptical orbits.
As a result, one should expect establishment of the virial
relation~\citep{Landau_76}:
\begin{equation}
\langle k \rangle = \frac{1}{2} \, | \langle u \rangle | \, ,
\label{eq:virial_theorem}
\end{equation}
where $ \langle k \rangle $ and $ \langle u \rangle $~are the average kinetic
and potential energies per one particle.
Strictly speaking, the virial relation involves time-averaged energies;
but we shall employ here the conjecture of ergodicity, \textit{i.e.},
to assume that the physical quantities averaged over time are equal to
the ones averaged over ensemble.

Next, the ensemble-averaged potential (Coulomb) energy can be estimated just
from the distance between the particles:
\begin{equation}
\langle u \rangle = C_u \frac{e^2}{\langle r \rangle} \, ,
\label{eq:avr_pot_en}
\end{equation}
where $ C_u $~is a numerical coefficient on the order of unity.
The average electron--ion distance, appearing in this formula,
can be evidently written as
\begin{equation}
\langle r \rangle = C_r \frac{1}{N^{1/3}} \, ,
\label{eq:avr_interpart_dist}
\end{equation}
where
$ N $~is the concentration of the charged particles of each kind, and
$ C_r $~is yet another numerical coefficient close to~1/2.

At last, the average kinetic energy, appearing in
formula~(\ref{eq:virial_theorem}), can be expressed through the effective
``virial'' temperature as
\begin{equation}
\langle k \rangle = \frac{3}{2} \, k_{\rm B} T_{\rm vir} \, .
\label{eq:avr_kin_en}
\end{equation}
At the first glance, this formula seems to be suitable only for an ideal
monatomic gas.
However, it is applicable actually at any strength of the Coulomb interaction.

Really, the multi-body partition function for the entire system can be
written in the most general form as
\begin{eqnarray}
&&
f(\textbf{r}_1, \dots , \textbf{r}_{2N},
  \textbf{v}_1, \dots , \textbf{v}_{2N}) =
\\
&&
\qquad
A_f \exp \bigg\{\!{-}\frac{1}{k_{\rm B} T_{\rm vir}}
  \Big[ \sum_{i=1}^{2N} \, \frac{ m_i \textbf{v}_i^2 }{2}
  + U( \textbf{r}_1, \dots , \textbf{r}_{2N}) \Big] \bigg\} ,
\nonumber
\label{eq:gen_distr_fun}
\end{eqnarray}
where
$ \textbf{r}_i $ and $ \textbf{v}_i $~are the coordinates and velocities of
all particles (both electrons and ions),
$ m_i $~are their masses,
$ N $~is the number of particles of each kind in the entire system or in
a unitary volume, and
$ U $~is the potential (Coulomb) energy of interaction between the charged
particles.

Even if the potential energy is a complex function of the coordinates and
cannot be treated as a small correction to the kinetic energy, an average
value of any physical quantity~$ G $ depending only on the velocity of
the $j$th~particle (\textit{e.g.}, its kinetic energy) can be calculated
exactly:
\begin{equation}
\langle G(\textbf{v}_j \rangle =
\frac{\displaystyle \int\! G(\textbf{v}_j)
  \exp \bigg\{ \!\! - \! \frac{1}{k_{\rm B} T_{\rm vir}}
  \frac{ m_j \textbf{v}_j^2 }{2} \bigg\} \, d \textbf{v}_j }%
  {\displaystyle %
  \int \! \exp \bigg\{ \!\! - \! \frac{1}{k_{\rm B} T_{\rm vir}}
  \frac{ m_j \textbf{v}_j^2 }{ 2 } \bigg\} \, d \textbf{v}_j } \; ,
\label{eq:gen_avr_proc}
\end{equation}
Really, integrals over the coordinates
$ \textbf{r}_1, \dots , \textbf{r}_{2N} $ in the numerator and denominator
are identical and automatically cancel each other at any strength of
the Coulomb interaction~$ U $.
On the other hand, a multi-dimensional integral over the velocities is
factorized into a product of integrals over each velocity, and they also
compensate each other for any velocity except of~$ \textbf{v}_j $.
So, taking the function~$ G $ to be a kinetic energy of the $j$th particle,
$ m_j \textbf{v}_j^2 / 2 $, we get just the above-written
relation~(\ref{eq:avr_kin_en}).

Finally, combining formulas~(\ref{eq:virial_theorem})--(\ref{eq:avr_kin_en}),
we obtain the virial temperature as function of the concentration of charged
particles:
\begin{equation}
k_{\rm B} T_{\rm vir} = \frac{1}{3} \frac{C_u}{C_r} \, e^2 N^{1/3} \, .
\label{eq:T_vir}
\end{equation}
Let us emphasize that, unfortunately, it would be impossible to relate
the effective temperature with the parameters of Coulomb interaction in
a more straightforward way, namely, by averaging the potential
energy~$ \langle u \rangle $ with the partition
function~(\ref{eq:gen_distr_fun}).
Really, because of the complex form of function~$ U $, it is hardly
possible to evaluate such integral in a reasonable way.
So, to get around this obstacle, we followed the indirect route:
to calculate exactly the average kinetic energy and then to relate it to
the potential one by the virial theorem, which is valid at any intensity of
the Coulomb interaction.

\begin{figure}[b]
\includegraphics[width=0.8\columnwidth]{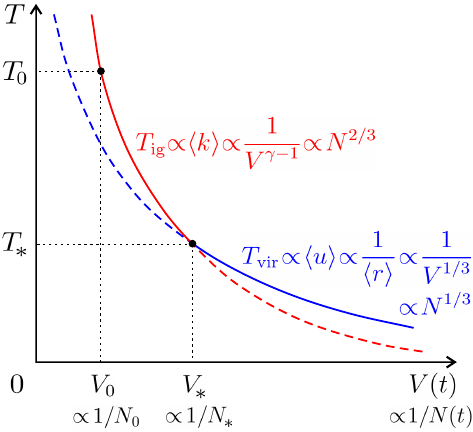}
\caption{\label{fig:Transition}
Temperature evolution in the ideal gas ($ T_{\rm ig} $, red curve) and in
the system of ``virialized'' charged particles ($ T_{\rm vir} $, blue curve).
The physically realized parts of these curves are shown by solid lines;
and the unphysical parts, by the dashed lines.
All physical quantities at the instant of transition between two regimes are
marked by asterisks.}
\end{figure}

The resulting expression~(\ref{eq:T_vir}), where temperature depends only
on the concentration, might look a bit strange.
However, it is well supported by the experimental finding~\cite{Fletcher_07}
that at the late stage of UCP expansion (when
$ N \propto 1 / V \propto t^{-3} $) the electronic temperature adapts just to
the variation of potential energy, \textit{i.e.}, decays as
$ T_e \propto N^{1/3} \propto t^{-1} $ instead of
$ T_e \propto N^{\gamma - 1} \propto t^{-2} $, as should be expected for
the adiabatic expansion of an ideal monatomic gas with
$ \gamma = 5/3 $.
Moreover, the variation with $ T_e \propto N^{1/3} $ formally corresponds to
$ \gamma = 4/3 $; and just this effective adiabatic index was predicted
earlier in paper~\cite{Zelener_04} for the case when the Coulomb
coupling parameter approaches a value about unity,
$ {\Gamma}_e{\approx}\,|\langle u \rangle| / \langle k \rangle\,{\sim}\,1 $.
The corresponding behavior is schematically illustrated in
Fig.~\ref{fig:Transition}.

Next, using the above-written expression for the virial temperature,
general formula~(\ref{eq:F-init}) for the one-particle distribution function
in the radius--energy variables can be rewritten in a more detailed form as
\begin{equation}
F(r, \varepsilon) =
  A_F \exp\bigg\{{-}\frac{3 \, (C_r/C_u)}{e^2 N^{1/3}} \, \varepsilon\bigg\}
  \frac{1}{\sqrt{\varepsilon - U_{\rm eff}(r)}} \, ,
\label{eq:F-fin}
\end{equation}
where the effective potential energy~$ U_{\rm eff} $ is given by
formula~(\ref{eq:Eff_pot}).

A crucial parameter in this ``centrifugal'' potential is the electronic
angular momentum~$ M_e $ with respect to the nearest ion, which is
approximately conserved because of adiabatic invariance.
In the case of a plasma bunch experiencing a free expansion,
this quantity can be estimated at the instant
of plasma transition to the strongly-coupled state.
(From here on, the corresponding quantities will be marked by asterisks.)
According to formula~(\ref{eq:T_vir}),
$ v_*{\propto}\,e \, m_e^{-1/2} N_*^{1/6} \! $,
while $ r_*{\propto}\,N_*^{-1/3} $.
Since $ M_e \propto m_e r_* v_* $, one can write:
\begin{equation}
M_e = C_M e \, m_e^{1/2} N_*^{-1/6} ,
\label{eq:ang_mom}
\end{equation}
where
$ C_M $~is a numerical factor on the order of unity.

At last, the quantity~$ N_* $, which is a crucial parameter in all subsequent
formulas, can be derived by the following way.
If $ N_0 $ and $ T_0 $~are the concentration and temperature at the initial
instant of time, then the law of evolution of the ideal gas (solid red curve
in Fig.~\ref{fig:Transition}) results in
$
T_* = T_0 (N_* / N_0)^{2/3} .
$
On the other hand, the same temperature~$ T_* $ at the end of this stage
can be estimated by formula~(\ref{eq:T_vir}) for the strongly-coupled state,
where all physical quantities should be written with asterisks.
So, combining these two formulas, we can easily get:
\begin{equation}
N_* = \frac{(C_u / C_r)^3 e^6 N_0^2}{27 (k_{\rm B} T_0)^3} \, .
\label{eq:N*_orig}
\end{equation}

\subsection{Dimensionless Variables}
\label{sec:Dimless}

To calculate the integrals~(\ref{eq:Nf_and_Nb-gen_def}), it is convenient
to rewrite all formulas in the dimensionless variables.
Let the unit of distance be~$ l / 2 $, where
\begin{equation}
l = N_*^{-1/3} .
\label{eq:l-def}
\end{equation}
(Since $ l $~is a typical separation between two particles of the same kind,
$ l / 2 $ should be the characteristic distance between the electron and ion.)
Then, the unit of time will be the Keplerian period of revolution of the electron
in an orbit with radius (or major semi-axis)~$ l / 2 $:
\begin{equation}
\tau = \frac{\pi}{\sqrt{2}} \frac{m_e^{1/2}}{e} \, l^{3/2} \, = \,
\frac{\pi}{\sqrt{2}} \, \frac{m_e^{1/2}}{e \, N_*^{1/3}} \, .
\label{eq:tau-def}
\end{equation}
So, the dimensionless radius and time, which will be marked by tildes,
are defined as
\begin{equation}
\tilde{r} =  2 r / l
\quad \mbox{and} \quad
\tilde{t} = t / \tau .
\label{eq:r_and_t-dimless}
\end{equation}
At last, the unit of energy can be obviously introduced as
\begin{equation}
\hat{\varepsilon} = 2 \, e^2 N_*^{1/3} ,
\label{eq:hat_eps-def}
\end{equation}
and all energies will be normalized to this quantity.

In the above-written units, the one-particle distribution
function~(\ref{eq:F-fin}), takes the form:
\begin{equation}
\tilde{F}(\tilde{r},\tilde{\varepsilon}) = \tilde{A}_F \,
  \frac{\displaystyle \exp \bigg\{ \!\! - 6 \, \frac{C_r}{C_u}
    \bigg(\!\frac{N}{N_*}\!\bigg)^{\!\! -1/3} \tilde{\varepsilon} \, \bigg\}}%
  {\displaystyle \bigg[ \tilde{\varepsilon} - \frac{C_M^2}{\tilde{r}^2}
    + \frac{1}{\tilde{r}}\bigg]^{1/2}} \, .
\label{eq:F-dimless}
\end{equation}
As shown in Fig.~\ref{fig:Phase_space}, the region of admissible states in
the electron phase space is $ \varepsilon \geqslant U_{\rm eff}(r) $ or,
in dimensionless units,
\begin{equation}
\tilde{\varepsilon} \geqslant \, \tilde{U}_{\rm eff}(\tilde{r}) =
  \frac{C_M^2}{\tilde{r}^2} - \frac{1}{\tilde{r}} \, .
\label{eq:admiss_reg}
\end{equation}

This region is subdivided by the horizontal
line~$ {\varepsilon}_0 = U_{\rm eff}(r_0) $ into two subregions,
$ {\rm D}_{\rm f} $ and $ {\rm D}_{\rm b} $, associated with free and bound
electronic states.
The corresponding boundary value reads as
\begin{equation}
\tilde{\varepsilon}_0 = \,
  \frac{C_M^2}{\tilde{r}_0^2} - \frac{1}{\tilde{r}_0} \, .
\label{eq:admiss_reg_bound}
\end{equation}
The boundary radius~$ r_0 $ of the potential well should be approximately
equal to half the distance between the ions:
\begin{equation}
r_0 = \frac{1}{2} \, C_0 N^{-1/3} ,
\label{eq:r_0-def}
\end{equation}
where $ C_0 $~is a numerical coefficient close to unity.
In the dimensionless form, it can be written as
\begin{equation}
\tilde{r}_0 = C_0 (N / N_*)^{-1/3} .
\label{eq:r_0-dimless}
\end{equation}
To avoid misunderstanding, let us emphasize that the coefficients~$ C_r $,
appearing in formula~(\ref{eq:avr_interpart_dist}), and~$ C_0 $, appearing
in~(\ref{eq:r_0-def}), are not identical: the first of them refers to
thermodynamic (or energetic) properties of the plasma, while the second
characterizes kinetic (or hopping) properties of the electrons.

\begin{figure}[t]
\includegraphics[width=0.85\columnwidth]{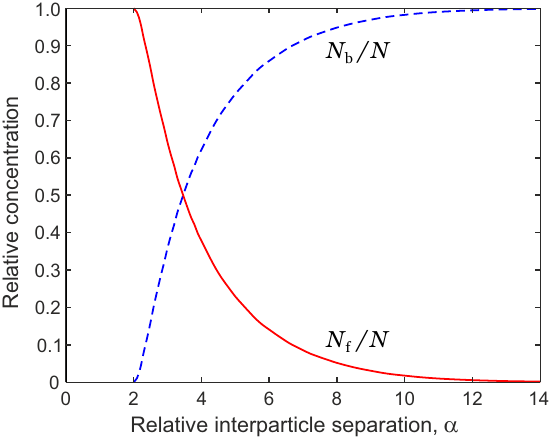}
\caption{\label{fig:Crossover_basic}
Relative concentration of the free, $ N_{\rm f} / N $, and
bound electrons, $ N_{\rm b} / N $, as function of the relative
interparticle separation $ \alpha = (N / N_*)^{-1/3} $ for the
``basic'' set of parameters ($ C_0 = C_M  = C_U = 1 $).}
\end{figure}

To write the final formulas, it is convenient to introduce yet another
dimensionless parameter:
\begin{equation}
\alpha = (N / N_*)^{-1/3} .
\label{eq:alpha-def}
\end{equation}
This is actually the relative interparticle separation, normalized to
its value at the instant of plasma transition into the strongly-coupled
state.
Besides, taking into account that the coefficients~$ C_r $ and~$ C_u $
appear in all formulas only as a ratio to each other and $ C_r \approx 1/2 $,
it is reasonable to introduce the new coefficient~$ C_U $ so that
\begin{equation}
C_r / C_u = 1 / (2 C_U) .
\label{eq:C_U-def}
\end{equation}
As a result, the one-particle distribution function~(\ref{eq:F-dimless}) will
take the form:
\begin{equation}
\tilde{F}(\tilde{r},\tilde{\varepsilon}) = \tilde{A}_F \,
  \frac{\displaystyle \exp \big\{ \!\! - \big( 3 / C_U \big) \alpha \,
    \tilde{\varepsilon} \big\}}%
  {\displaystyle \bigg[ \tilde{\varepsilon} - \frac{C_M^2}{\tilde{r}^2}
    + \frac{1}{\tilde{r}}\bigg]^{1/2}} \, ;
\label{eq:F-dimless-fin}
\end{equation}
and concentration of the charged particles at the instant of transition to
the strongly-coupled state~(\ref{eq:N*_orig}) can be rewritten as
\begin{equation}
N_* = \frac{8 C_U^3 e^6 N_0^2}{27 (k_{\rm B} T_0)^3} \, .
\label{eq:N*_modif}
\end{equation}

Let~$ I_{\rm f} $ be the integral of function~$ \tilde{F} $ over
the region~$ {\rm D}_{\rm f} $ depicted in Fig.~\ref{fig:Phase_space}; and
$ I_{\rm b} $, over the region~$ {\rm D}_{\rm b} $:
\begin{equation}
I_{\rm f} \! = \!
  \iint\limits_{{\rm D}_{\rm f}} \! \tilde{F}(\tilde{r}, \tilde{\varepsilon})
    \, d\tilde{r} \, d\tilde{\varepsilon}
\;\; \mbox{and} \;\;
I_{\rm b} \! = \!
  \iint\limits_{{\rm D}_{\rm b}} \! \tilde{F}(\tilde{r}, \tilde{\varepsilon})
    \, d\tilde{r} \, d\tilde{\varepsilon} \, .
\label{eq:If_and_Ib-gen}
\end{equation}
Then, relative concentrations of the free and bound electrons will be
evidently given by formulas:
\begin{equation}
\frac{N_{\rm f}}{N} = \frac{I_{\rm f}}{I_{\rm f} + I_{\rm b}}
\;\; \mbox{and} \;\;
\frac{N_{\rm b}}{N} = \frac{I_{\rm b}}{I_{\rm f} + I_{\rm b}} \, ,
\label{eq:Nf_and_Nb-rel}
\end{equation}
respectively.

As can be easily found, minimum of the function
$ \tilde{U}_{\rm eff}(\tilde{r}) $ specified by
expression~(\ref{eq:admiss_reg}) is achieved at
\begin{equation}
\tilde{r}_{\rm min} \! = 2 \, C_{\! M}^2
\label{eq:r_min}
\end{equation}
and equals
\begin{equation}
(\tilde{U}_{\rm eff} \!)_{\rm min} = - 1 / (4 \, C_{\! M}^2) \, .
\label{eq:U_min}
\end{equation}
Since the bound electron states can evidently exist only if
$ \, \tilde{r}_{\rm min} < \tilde{r}_0 $ (and, therefore,
$ (\tilde{U}_{\rm eff} \!)_{\rm min} < \tilde{\varepsilon}_0 $),
the integral~$ I_{\rm b} $ will be nonzero only at 
\begin{equation}
\alpha > \, 2 \, C_{\! M}^2 / C_0 \, .
\label{eq:Ib_nonzero}
\end{equation}

\section{Results and Discussion}
\label{sec:Results}

\begin{figure}[t]
\includegraphics[width=0.85\columnwidth]{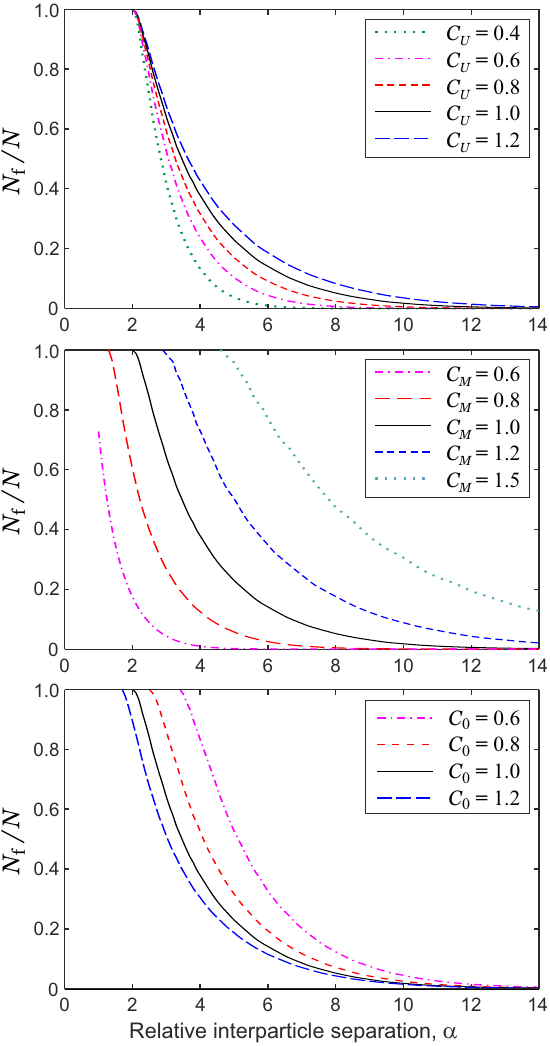}
\caption{\label{fig:Crossover_var_par}
Variations of the $ [N_{\rm f} / N] (\alpha) $ profiles depending
on the coefficients $ C_U $. $ C_M $, and $ C_0 $, when two other
coefficients are fixed at unity.}
\end{figure}

The integrals~(\ref{eq:If_and_Ib-gen}) appearing in
the expressions~(\ref{eq:Nf_and_Nb-rel}) can be calculated numerically.
Since our model involves a few semi-empirical parameters (namely, $ C_0 $,
$ C_M $, and $ C_U $), we should consider also how sensitive the required
quantities~$ N_{\rm f} / N $ and $ N_{\rm b} / N $ are to these parameters.

The ``basic'' case, when all parameters are equal to unity, is presented in
Fig.~\ref{fig:Crossover_basic}.
We can see here that the relative concentration of free
electrons~$ N_{\rm f} / N $ sharply drops, while the relative concentration of
bound electrons~$ N_{\rm b} / N $ sharply jumps with increase in the relative
interparticle separation~$ \alpha $.
The critical value of this crossover turns out
to be~$ {\alpha}_{\rm cr} \! \approx 3 $, which closely resembles Mott
transition in the condensed-matter physics.
However, such a coincidence is mostly occasional, because the corresponding
models are based on the substantially different assumptions.

Next, Fig.~\ref{fig:Crossover_var_par} demonstrates how the Rydberg
gas--plasma crossover depends on the parameters involved in our model.
It is seen that the curves $ [N_{\rm f} / N] (\alpha) $ are relatively
insensitive to the variations of~$ C_U $ and $ C_0 $ but depend substantially
on~$ C_M $.
So, it would be desirable in future to specify this parameter in a more
robust way.

Let us mention that yet another example of collective effects in the dynamics
of Rydberg atoms is the crystalline-like Rydberg matter;
\textit{e.g.}, paper~\cite{Manykin_94} and references therein.
However, this is irrelevant to the topic of the present study, which is
devoted to the usual disordered gases and plasmas.

In summary, we presented a theoretical framework for the description of
collective ionization--recombination balance in a mixture of ultracold
neutral and ionized Rydberg atoms.
This should be a promising direction for further research in
the recently-started experiments with steady-state ultracold
plasmas~\cite{Zelener_24}.

\acknowledgments

Yu.\,V.~Dumin is grateful to
E.\,R.~Grant,
A.\,M.~Ignatov,
V.\,S.~Vorob'ev,
S.~Whitlock, and
D.\,I.~Zhukhovitskii
for fruitful discussions and valuable suggestions.

\subsection*{Funding}

This work was supported by the ongoing institutional funding.
No additional grants to carry out or direct this particular research were
obtained.

\subsection*{Conflict of Interests}

The authors declare no conflict of interests.


\end{document}